\documentclass[twocolumn,showpacs,amsmath,amssymb]{revtex4}
\usepackage{graphicx}
\usepackage{dcolumn}
\usepackage{bm}
\usepackage{slashed}
\usepackage{amsmath}
\usepackage{float}
\usepackage{subfig}

\def\be{\begin{equation}}
\def\ee{\end{equation}}
\def\ba{\begin{array}}
\def\ea{\end{array}}
\def\bea{\begin{eqnarray}}
\def\eea{\end{eqnarray}}

\begin{document}
\title{Placing the newly observed state $B_{J}(5840)$ in  bottom spectra
 along with states $B_{1}(5721)$, $B_{2}^{*}(5747)$, $B_{s1}(5830)$, $B_{2s}^{*}(5840)$
 and $B_{J}(5970)$
}
\author{\large  Pallavi Gupta}
\author{\large A. Upadhyay}
\affiliation{School of Physics and Materials Science, Thapar
Institute of engineering and Technology, Patiala - 147004, Punjab,
INDIA}
\date{\today}
\baselineskip =1\baselineskip

\begin{abstract}
In this article, we apply the formalism of Ref.\cite{17} to discuss
the quantum number assignments for recently observed $B_{J}(5840)$
state by LHCb collaboration \cite{9}, we classify the six possible
$J^{P}$'s for this state on the basis of the theoretically available
masses. By analyzing the strong decay widths and the branching
ratios for all these six cases of $B_{J}(5840)$, we justify one of
them to be the most favorable assignment for it. We also examined
the recently observed bottom state $B_{J}(5970)$ as 2S$1^{-}$ and
states $B_{J}(5721)$ and $B_{2}^{*}(5747)$ with their strange
partners $B_{s1}(5830)$ and $B_{2s}^{*}(5840)$ for their $J^{P}$'s
as $1P_{3/2}1^{+}$ and $1P_{3/2}2^{+}$ respectively. The predicted
coupling constants $g_{XH}$, $\widetilde{g}_{HH}$ and $g_{TH}$ helps
in redeeming the strong decay width of experimentally missing bottom
states $B(2 ^{1}S_{0})$, $B_{s}(2 ^{3}S_{1})$, $B_{s}(2 ^{1}S_{0})$,
$B(1 ^{1}D_{2})$, $B_{s}(1 ^{3}D_{1})$ and $B_{s}(1 ^{1}D_{2})$.
These predictions provide a crucial information for upcoming
experimental studies.
\end{abstract}

\pacs{ 12.39.Hg, 13.25.Hw, 14.40.Nd} \keywords{Heavy quark effective
theory,Decays of bottom mesons, Bottom mesons $(|B|>0)$ }
 \maketitle

\section{Introduction}

In the recent decades, a significant experimental progress have been
achieved in studying the heavy-light meson spectroscopy. Heavy-light
mesons composed of one heavy quark Q and a light quark
$\overline{q}$ are useful in understanding the strong interactions
in the non perturbative regime. Recently, many new charm states like
$D^{*}_{2}(3000)$, $D_{J}(3000)$, $D_{J}^{*}(3000)$,
$D^{*}_{3}(2760)$, $D_{1}^{*}(2680)$, $D^{*}_{2}(2460)$,
$D^{*}_{J}(2650)^{0}$, $D^{*}_{J}(2760)^{0}$ etc announced by LHCb
\cite{1,2} and BABAR \cite{3} have successfully stimulated the charm
meson spectroscopy. Whereas in bottom sector, only ground state
$B^{0}(5279)$, $B^{\pm}(5279)$, $B^{*}(5324)$, $B_{s}(5366)$,
$B^{*}_{s}(5415)$ and few of the low lying excited bottom mesons
$B_{1}(5721)$, $B^{*}_{2}(5747)$
 are experimentally well known
\cite{4,5,6,7,10,11} which are listed in PDG \cite{8}. But the
information for other excited bottom mesons is rather limited as
compared to the charm mesons. However, the recent measurement of
newly observed bottom mesons by LHCb have opened the gate to extend
our understanding for these higher excited bottom states.\\ Recently
in 2015, LHCb has reported the observation of $B_{J}(5721)^{0,+}$
and $B_{2}^{*}(5747)^{0,+}$ states, along with the observation of
two new resonances $B_{J}(5840)^{0,+}$ and $B_{J}(5960)^{0,+}$ in
the pp collision data, at center-of-mass energies of 7 and 8 TeV
\cite{9}. Also in 2013, the CDF collaboration analyzed a new state
$B_{J}(5960)$ both in the $B^{0}\pi^{+}$ and $B^{+}\pi^{-}$ mass
distribution from the p$\overline{p}$ collision data at $\sqrt{s}$ =
1.96 TeV \cite{12}.

And in the strange sector of the bottom mesons, $B_{1s}(5830)$ and
$B_{2s}^{*}(5840)$ states are well observed by CDF \cite{38,7},
 D0\cite{39} and LHCb \cite{37} collaborations and are assigned the $J^{P}$ states $1P_{3/2}1^{+}$ and $1P_{3/2}2^{+}$ respectively.
 The masses and the widths of the recently measured experimental bottom states  $B_{J}(5721)$,
 $B_{2}^{*}(5747)$, $B_{J}(5840)$, $B_{J}(5960)$, $B_{sJ}(5830)$ and $B_{2s}^{*}(5840)$ are listed in
 Table~\ref{tab1}. Assigning a place in the mass spectra for such newly observed experimental states
is very important, as the $J^{P}$'s helps in redeeming many crucial
strong interaction properties of the states. To assign a $J^{P}$,
many theoretical models are available such as quark model
\cite{13,14,15,41b}, Heavy Quark Effective Theory (HQET)\cite{17},
$^{3}P_{0}$ model \cite{16,41b} and many
 more \cite{18}. Many theoretical
predictions have been made for assigning a
 particular $J^{P}$'s to these newly observed states. Different theoretical approaches uses
  different theoretical
  parameters and therefore the predictions are not
 completely consistent with each other, hence a particular $J^{P}$
 is not confirmed for these experimentally observed bottom states.

\setlength{\tabcolsep}{0.09em} %
{\renewcommand{\arraystretch}{0.2}%
\begin{table*}{\normalsize
\renewcommand{\arraystretch}{1.0}
\tabcolsep 0.2cm \caption {\label{tab1} Values of the masses and the
decay widths of bottom mesons observed by various collaborations.}
\footnotesize
\begin{tabular*}{170mm}{@{\extracolsep{\fill}}cccccc}
\toprule State&$J^{P}$&Mass/MeV&Width/MeV& Experiments & Observed Decay Modes\\
\hline $B_{J}(5721)$&$1^{+}$&$5727.7\pm0.7$&$30.1\pm 1.5$&LHCb
\cite{9}&$B^{*}\pi$\\
&&$5720.6\pm2.4$&-&D0\cite{10}&$B^{*}\pi$\\
&&$5725.3\pm1.6$&-&CDF\cite{11}&$B^{*}\pi$\\
\hline $B_{2}^{*}(5747)$&$2^{+}$&$5739.44\pm0.37$&$24.5\pm1.0$&LHCb
\cite{9}&$B^{*}\pi$,$B\pi$\\
&&$5746.8\pm2.4$&-&D0\cite{10}&$B^{*}\pi$,$B\pi$\\
&&$5740.2\pm1.7$&$22.7\pm3.2$&CDF\cite{11}&$B^{*}\pi$,$B\pi$\\
\hline $B_{J}(5840)$&$-$&$5862.9\pm5.0$&$127.4\pm16.7$&LHCb
\cite{9}&$B\pi$\\
\hline $B_{J}(5960)$&$-$&$5978\pm5$&-&CDF\cite{11}&$B\pi$\\
&&$5969.2\pm2.9$&$82.3\pm7.7$&LHCb
\cite{9}&$B\pi$\\
\hline $B_{s1}(5830)$&$1^{+}$&$5828.40\pm0.04$&-&LHCb\cite{37}&$B^{*}K$\\
&&$5828.3\pm0.1$&$0.5\pm0.3$&CDF\cite{38}&$B^{*}K$\\
&&$5829.4\pm0.7$&-&CDF\cite{7}&$B^{*}K$\\
\hline $B_{s2}^{*}(5840)$&$2^{+}$&$5839.6\pm1.1$&-&D0\cite{39}&$B^{*}K$,BK\\
&&$5839.70\pm0.7$&-&CDF\cite{7}&$B^{*}K$,BK\\
&&$5839.70\pm0.1$&$1.40\pm0.4$&CDF\cite{38}&$B^{*}K$,BK\\
&&$5839.99\pm0.05$&$1.56\pm0.13$&Lhcb\cite{37}&$B^{*}K$,BK\\ \hline
\end{tabular*}}
\end{table*}

The first two bottom states in Table~\ref{tab1} $B_{J}(5721)$ and
$B_{2}^{*}(5747)$ are being analyzed theoretically by various models
\cite{19,20,21,22,41b} and their analysis have interpreted the
$B_{2}^{*}(5747)$ state to belong to $J^{P}$ $2^{+}$. For the
$B_{J}(5721)$ state, some of the theoretical works \cite{21,17}
favors it to be the spin partner of the $B_{2}^{*}(5747)$ state and
hence $J^{P}$ as $1^{+}$ for $j_{l}=3/2$ P-wave bottom meson. And
other papers \cite{19,20,22} suggests $B_{J}(5721)$ to be the
mixture of the $1P_{1/2}$ and $1P_{3/2}$ state. The other two bottom
states $B_{1s}(5830)$ and $B_{2s}^{*}(5840)$,
 being the strange partners of $B_{J}(5721)$ and $B_{2}^{*}(5747)$ states, too belongs to
$1P_{s3/2} 1^{+}$ and $1P_{s3/2} 2^{+}$ $J^{P}$'s respectively.
\\ For the bottom state $B_{J}(5960)$, authors in Ref. \cite{23} claimed, that
 the properties of the $B(5970)$ seen by CDF collaboration \cite{12} are consistent
 with the properties of the $B_{J}(5960)$ measured by LHCb \cite{9}, so they may be the same state.
The theoretical analysis made by studying decay widths for
$B_{J}(5960)$ by QPC model \cite{22} and HQET \cite{22a}, favors it
to belong to 2S$1^{-}$ state. This prediction is also supported by
work in \cite{24} , where the authors have used the relativistic
quark model. In our previous work \cite{22b}, where masses were
predicted using the QCD and $1/m_{Q}$ corrections to the flavor
independent parameters $\Delta_{F}$ and $\lambda_{F}$, $B_{J}(5960)$
is again favored to be the 2S$1^{-}$ state. But, Qi-Fang Lu et. al.
in Ref.\cite{20} studied masses and strong decays of
 $B_{J}(5960)$ states with different spin parity hypothesis and
 identified that the $B_{J}(5960)$ belongs to
 1D$3^{-}$ state. A review of
 the open charm and open bottom systems by Hua-Xing Chen in
 Ref.\cite{25} undergoes various theoretical analysis with
 conclusion that $B_{J}(5960)$ belongs to 2S$1^{-}$ state.
\\As most of the analysis favored the $2S1^{-}$ spin parity, thus
$B_{J}(5960)$ is considered to be the radial excited $2S1^{-}$
state. And lastly for the $B_{J}(5840)$ bottom state, two spin
parity proposals has been put forward. First one is given by authors
in \cite{20}, where they suggested it to belong to 2S$0^{-}$ state.
This interpretation matches with the LHCb collaboration analysis
\cite{9}. Second possible $J^{P}$ is given in \cite{24}, where
authors suggested the $B_{J}(5840)$ state to be the member of
1P$1^{+}$ doublet with $j_{l}=1/2$. As $B_{J}(5840)$ is studied only
in few papers, so in this scenario, $B_{J}(5840)$ needs to be
properly placed in the bottom meson spectra. In Ref. \cite{20}, the
$J^{P}$ for bottom state $B_{J}(5840)$ has been analysed by
predicting the masses and decay widths using non-relativistic quark
model and $^{3}P_{0}$ model respectively. Whereas the $J^{P}$  in
Ref. \cite{24} has been decided just on the basis of theoretically
predicted bottom meson masses. In both the references, the models
have some unknown parameters, which are fitted by using experimental
data's like decay width of bottom state $B_{2}^{*}(5747)$. So the
accuracy of these predictions cannot be completely justified.
\\We apply the Heavy Quark Effective Theory (HQET)
 to  discuss the assignments of the quantum
  numbers $J^P$ of the
open bottom states recently reported by LHCb \cite{9}. In the past
decades, HQET have successfully explained the properties of
heavy-light hadrons. The effectiveness of this theory lies on the
fact that, heavy quark is treated as a dynamically degree of
freedom. As a result, number of unknown parameters is greatly
reduced by using the heavy quark spin and flavor symmetry. Another
peculiar property of HQET is that each effective quark field is
written in terms of four vector $v_{\mu}$ of heavy quark, which
remains conserved in the strong interactions in the infinitely heavy
quark mass limit. In our paper, we predicted $J^{P}$ using the
branching ratio $\frac{B\pi}{B^{*}\pi}$ which is free from any
theoretical parameter, hence the prediction made by HQET are
supposed to be more accurate and logical. The HQET was originally
proposed to study the interactions among heavy-light charm and
bottom mesons through the emission of light pseudo-scalar mesons
($\pi,\eta,K$)\cite{25a,25b,25c,25d,25e,25f,25g,25h,25i,33}. The
paper is arranged as follows: section 2 gives the description about
the model "Heavy Quark Effective Theory". Section 3 represents the
numerical analysis where we investigated the $B_{J}(5840)$ state by
considering all the possible quark model assignments based on its
mass and spin parity and analyze the
 branching ratio $\frac{B\pi}{B^{*}\pi}$ for all these possible $J^{P}$ states and
 confirm one of them to be the suitable $J^{P}$ for this
 state. Along with this, we also analyze
the bottom states $B_{1}(5721)$, $B_{2}^{*}(5747)$, $B_{1s}(5830)$,
$B_{2s}^{*}(5840)$ and $B_{J}(5960)$ for their respective $J^{P}$'s.
In addition to this, we also study the strong decays for the
experimentally unobserved but theoretically predicted states $B(2
^{1}S_{0})$, $B_{s}(2 ^{3}S_{1})$, $B_{s}(2 ^{1}S_{0})$, $B(1
^{1}D_{2})$, $B_{s}(1 ^{3}D_{1})$ and $B_{s}(1 ^{1}D_{2})$  and
discuss their strong coupling constants involved. Section 4 presents
the summary of our work.

\section{Framework}

We use  heavy quark effective theory for the study of heavy light
bottom mesons. As in our analysis we use the decay width expressions
calculated in Ref. \cite{17} by means of HQET, it is useful to
remind the theoretical framework of this theory. In the heavy quark
limit $m_{Q}\rightarrow\infty$,(Q=c,b), spin of
 the heavy quark $s_{Q}$ decouples from the light degree of freedom
 which includes light anti-quark and the gluons. Thus the spin of the heavy quark
$s_{Q}$ and the total angular momentum of light degree of freedom
$s_{l}$ are separately conserved. The total angular momentum of
light degree of freedom $s_{l}$ is given by
($s_{l}=s_{\overline{q}}+l$) where, $s_{\overline{q}}$ is the spin
of the light quark and l is the orbital angular momentum of the
light degree of freedom, therefore the resultant angular momentum J
for each heavy-light meson is $J=s_{l} + s_{Q}$. Thus for each
state, there is a degenerate doublet of meson state with
$J^{P}=s_{l}^{P}\pm1/2$ which for S-wave $(\textit{l} =0)$, gives
the doublet which is represented by $(P, P^{*})$ with
$J^{P}_{s_{l}}= (0^{-},1^{-})_{\frac{1}{2}}$. For the P-wave
$(\textit{l} =1)$, we get two doublets which are represented by
$(P^{*}_{0},P^{'}_{1})$ and $(P_{1},P^{*}_{2})$ with
$J^{P}_{s_{l}}=(0^{+},1^{+})_{\frac{1}{2}}$ and
$(1^{+},2^{+})_{\frac{3}{2}}$ respectively. Similarly, two doublets
for the D-wave ($\textit{l}=2$)  are represented by
$(P^{*}_{1},P_{2})$ and $(P_{2}^{'},P^{*}_{3})$ belonging to
$J^{P}_{s_{l}}=(1^{-},2^{-})_{\frac{3}{2}}$ and
$(2^{-},3^{-})_{\frac{5}{2}}$ respectively. The above mentioned
doublets for each wave are expressed by the effective super-field
$H_{a}, T_{a}, X_{a}$ and $Y_{a}$ \cite{28,33}.
\begin{eqnarray}
\label{eq1} H_{a}=\frac{1+\slashed
v}{2}\{P^{*}_{a\mu}\gamma^{\mu}-P_{a}\gamma_{5}\}.
\end{eqnarray}
\begin{eqnarray}
T^{\mu}_{a}=\frac{1+\slashed v}{2}
\{P^{*\mu\nu}_{2a}\gamma_{\nu}-P_{1a\nu}\sqrt{\frac{3}{2}}\gamma_{5}
[g^{\mu\nu}-\frac{\gamma^{\nu}(\gamma^{\mu}-\upsilon^{\mu})}{3}]\}.
\end{eqnarray}
\begin{eqnarray}
X^{\mu}_{a}=\frac{1+\slashed
v}{2}\{P^{\mu\nu}_{2a}\gamma_{5}\gamma_{\nu}-P^{*}_{1a\nu}\sqrt{\frac{3}{2}}[g^{\mu\nu}\nonumber\\
-\frac{\gamma_{\nu}(\gamma^{\mu}+v^{\mu})}{3}]\}.
\end{eqnarray}
\begin{eqnarray}
Y^{\mu\nu}_{a}=\frac{1+\slashed
v}{2}\{P^{*\mu\nu\sigma}_{3a}\gamma_{\sigma}-P^{\alpha\beta}_{2a}\sqrt{\frac{5}{3}}\gamma_{5}[g^{\mu}_{\alpha}g^{\nu}_{\beta}\nonumber\\
-\frac{g^{\nu}_{\beta}\gamma_{\alpha}(\gamma^{\mu}-v^{\mu})}{5}-\frac{g^{\mu}_{\alpha}\gamma_{\beta}(\gamma^{\nu}-v^{\nu})}{5}]\}.
\end{eqnarray}
where the field $H_{a}$ describe the doublet of S-wave and field
$T_{a}$ represents the P-wave doublet $(1^{+},2^{+})_{\frac{3}{2}}$.
D-wave doublets are represented by $X_{a}$ and $Y_{a}$ fields for
the $(1^{-},2^{-})_{\frac{3}{2}}$ and $(2^{-},3^{-})_{\frac{5}{2}}$
$J^{P'}s$ respectively. Here indices a or b in the subsequent fields
are $SU(3)$ flavor index (u, d or s). The heavy meson field
$P^{(*)}$ contain a factor $\sqrt{m_{Q}}$ with mass dimension of
$\frac{1}{2}$. For the radially excited states for radial quantum
number n=2, these states are replaced by notation with $\sim$ on
their heads e.g. $\widetilde{P}, \widetilde{P}^{*}$ and so on. The
strong interaction for these heavy-light mesons involves their decay
with the emission of  light pseudoscalar mesons($\pi, \eta, K$),
which can be studied with the help of chiral perturbation theory.

 The light pseudoscalar mesons are
described by the fields $\xi= exp^{\frac{i\mathcal{M}}{f_{\pi}}}$,
where $\mathcal{M}$ is defined as
\begin{center}
\begin{eqnarray}
\mathcal{M} = \begin{pmatrix}
\frac{1}{\sqrt{2}}\pi^{0}+\frac{1}{\sqrt{6}}\eta & \pi^{+} & K^{+}\\
\pi^{-} & -\frac{1}{\sqrt{2}}\pi^{0}+\frac{1}{\sqrt{6}}\eta &
K^{0}\\
K^{-} & \overline{K}^{0} & -\sqrt{\frac{2}{3}}\eta
\end{pmatrix}.
\end{eqnarray}
\end{center}
The pion octet is introduced by the vector and axial vector
combinations
$V^{\mu}=\frac{1}{2}(\xi\partial^{\mu}\xi^{\dag}+\xi^{\dag}\partial^{\mu}\xi)$
and
$A^{\mu}=\frac{1}{2}(\xi\partial^{\mu}\xi^{\dag}-\xi^{\dag}\partial^{\mu}\xi)$.
We choose $f_{\pi}=130MeV$. The Dirac structure of the chiral
Lagrangian is given by the velocity vector v/c. The interaction
terms between the ground state doublet ($H_{a}$) and the excited
states ($T_{a}, X_{a}, Y_{a}$) through light pseudoscalar mesons are
written as :
 \begin{center}
\begin{eqnarray}
\label{eq:lagrangian} L_{HH}=g_{HH}Tr\{\overline{H}_{a}
H_{b}\gamma_{\mu}\gamma_{5}A^{\mu}_{ba}\}.\\
L_{TH}=\frac{g_{TH}}{\Lambda}Tr\{\overline{H}_{a}T^{\mu}_{b}(iD_{\mu}\slashed
A + i\slashed D A_{\mu})_{ba}\gamma_{5}\}+h.c..\\
L_{XH}=\frac{g_{XH}}{\Lambda}Tr\{\overline{H}_{a}X^{\mu}_{b}(iD_{\mu}\slashed
A + i\slashed D A_{\mu})_{ba}\gamma_{5}\}+h.c..
\end{eqnarray}
\begin{eqnarray}
L_{YH}=\frac{1}{\Lambda^{2}}Tr\{\overline{H}_{a}Y^{\mu\nu}_{b}[k^{Y}_{1}\{D_{\mu}
,D_{\nu}\}A_{\lambda}+k^{Y}_{2}(D_{\mu}D_{\lambda}A_{\nu}\nonumber\\
+D_{\nu}D_{\lambda}A_{\mu})]_{ba}\gamma^{\lambda}\gamma_{5}\}+h.c..
\end{eqnarray}

\end{center}
In these equations  \begin{eqnarray} D_{\mu} =
\partial_{\mu}+V_{\mu}, \hspace{0.5cm} \& \hspace{0.5cm}  \{D_{\mu},D_{\nu}\}
= D_{\mu}D_{\nu}+D_{\nu}D_{\mu},\nonumber\\
  \{D_{\mu} ,D_{\nu}D_{\rho}\}
=
D_{\mu}D_{\nu}D_{\rho}+D_{\mu}D_{\rho}D_{\nu}+D_{\nu}D_{\mu}D_{\rho}\nonumber\\+D_{\nu}D_{\rho}D_{\mu}+D_{\rho}D_{\mu}
D_{\nu}+D_{\rho}D_{\nu}D_{\mu} \end{eqnarray}

 Here $g_{HH}$, $g_{TH}$, $g_{XH}$ and $g_{YH} = k^{Y}_{1}+k^{Y}_{2}$
are the strong coupling constants, $\Lambda$ is the chiral symmetry
breaking scale which is taken as 1 GeV. Using the lagrangian
$L_{HH},L_{TH},L_{XH}$ and $L_{YH}$
 the two body strong decay widths of $Q\bar{q}$ heavy-light
 bottom mesons are calculated in Ref. \cite{17} as:
 \begin{equation} \Gamma=\frac{1}{(2J+1)} \sum \frac{p_M}
 {8 \pi M_i^2} |A|^2 \end{equation}
 where $A$ is the scattering amplitude, $p_{M}$ and $m_{M}$ are the final momentum and mass of the light
pseudo-scalar meson with $p_{M}$ =
$\sqrt{\lambda(M_{i}^{2},m_{M}^{2},M_{f}^{2})}/2M_{i}$, where
$\lambda(a,b,c) = a^{2}+b^{2}+c^{2}-2ab-2ac-2bc$ is the Källen
function. $M_{i}$ and $M_{f}$ stand for initial and final
heavy-meson mass.
\\$(0^{-},1^{-}) \rightarrow (0^{-},1^{-}) + M$
\begin{eqnarray}
\label{eq:lagrangian} \Gamma(1^{-} \rightarrow 1^{-})=
C_{M}\frac{g_{HH}^{2}M_{f}p_{M}^{3}}{3\pi f_{\pi}^{2}M_{i}}\\
\Gamma(1^{-} \rightarrow 0^{-})=
C_{M}\frac{g_{HH}^{2}M_{f}p_{M}^{3}}{6\pi f_{\pi}^{2}M_{i}}\\
\Gamma(0^{-} \rightarrow 1^{-})=
C_{M}\frac{g_{HH}^{2}M_{f}p_{M}^{3}}{2\pi f_{\pi}^{2}M_{i}}
\end{eqnarray}


 $(1^{+},2^{+}) \rightarrow (0^{-},1^{-}) + M$
\begin{eqnarray}
\label{eq:lagrangian} \Gamma(2^{+} \rightarrow 1^{-})=
C_{M}\frac{2g_{TH}^{2}M_{f}p_{M}^{5}}{5\pi f_{\pi}^{2}\Lambda^{2}M_{i}}\\
\Gamma(2^{+} \rightarrow 0^{-})=
C_{M}\frac{4g_{TH}^{2}M_{f}p_{M}^{5}}{15\pi f_{\pi}^{2}\Lambda^{2}M_{i}}\\
\Gamma(1^{+} \rightarrow 1^{-})=
C_{M}\frac{2g_{TH}^{2}M_{f}p_{M}^{5}}{3\pi
f_{\pi}^{2}\Lambda^{2}M_{i}}
\end{eqnarray}

$(1^{-},2^{-}) \rightarrow (0^{-},1^{-}) + M$
\begin{eqnarray}
\label{eq:lagrangian} \Gamma(1^{-} \rightarrow 0^{-})=
C_{M}\frac{4g_{XH}^{2}}{9\pi f_{\pi}^{2}\Lambda^{2}}
\frac{M_{f}}{M_{i}}[p_{M}^{3}(m_{M}^{2}+p_{M}^{2})]\\
\Gamma(1^{-} \rightarrow 1^{-})= C_{M}\frac{2g_{XH}^{2}}{9\pi
f_{\pi}^{2}\Lambda^{2}}
\frac{M_{f}}{M_{i}}[p_{M}^{3}(m_{M}^{2}+p_{M}^{2})]\\
\Gamma(2^{-} \rightarrow 1^{-})= C_{M}\frac{2g_{XH}^{2}}{3\pi
f_{\pi}^{2}\Lambda^{2}}
\frac{M_{f}}{M_{i}}[p_{M}^{3}(m_{M}^{2}+p_{M}^{2})]
\end{eqnarray}

$(2^{-},3^{-}) \rightarrow (0^{-},1^{-}) + M$

\begin{eqnarray}
\label{eq:lagrangian} \Gamma(2^{-} \rightarrow 1^{-})=
C_{M}\frac{4g_{YH}^{2}}{15\pi f_{\pi}^{2}\Lambda^{4}}
\frac{M_{f}}{M_{i}}[p_{M}^{7}]\\
\Gamma(3^{-} \rightarrow 0^{-})= C_{M}\frac{4g_{YH}^{2}}{35\pi
f_{\pi}^{2}\Lambda^{4}} \frac{M_{f}}{M_{i}}[p_{M}^{7}]\\
\Gamma(3^{-} \rightarrow 1^{-})= C_{M}\frac{16g_{YH}^{2}}{105\pi
f_{\pi}^{2}\Lambda^{4}} \frac{M_{f}}{M_{i}}[p_{M}^{7}]
\end{eqnarray}

In these equations, the coefficients $C_{\pi^{\pm}}, C_{K^{\pm}},
C_{K^{0}}, C_{\overline{K}^{0}}=1$, $C_{\pi^{0}}=\frac{1}{2}$ and
$C_{\eta}=\frac{2}{3}$ or $\frac{1}{6}$ as from Ref.\cite{17}.
Different values of $C_{\eta}$ corresponds to the initial state
being $b\overline{u}, b\overline{d}$ or $ b\overline{s}$
respectively. For the decay within n=1, the hadronic coupling
constants are notated as $g_{HH}$, $g_{TH}$ etc, and for the decay
from n=2 to n=1 these couplings are notated as $\widetilde{g}_{HH}$,
$\widetilde{g}_{TH}$ respectively.
 Higher order corrections for spin and
flavor violation of order $\frac{1}{m_{Q}}$ are excluded to avoid
new unknown coupling constants. The coupling constants involved in
these widths, can either be theoretically predicted or can be
determined indirectly from the known experimental values of the
decay widths. The numerical masses of various mesons used in the
calculation are listed in Table~\ref{input}

\setlength{\tabcolsep}{0.09em} %
{\renewcommand{\arraystretch}{0.2}%
\begin{table*}{\normalsize
\renewcommand{\arraystretch}{1.0}
\tabcolsep 0.2cm \caption{ \label{input}  Numerical value of the
meson masses used in this work \cite{8}.} \footnotesize
\begin{tabular*}{80mm}{c@{\extracolsep{\fill}}cccccc}
\toprule States &$B^{0}$&$B^{\pm}$&$B^{*}$&$B_{s}$&$B^{*}_{s}$\\
  \hline
  Masses/MeV&5279.58&5279.25&5325.20&5366.77&5415.40\\
  \hline
  States &$\pi^{\pm}$&$\pi^{0}$&$\eta$&$K^{+}$&$K^{0}$\\
  \hline
  Masses/MeV&139.57&134.97&547.85&493.67&497.61\\
  \hline
\end{tabular*}}
\end{table*}
\section{Numerical Analysis}

To assign a particular $J^{P}$ to the experimental available states
is very important, as the $J^{P}$'s helps in redeeming many crucial
strong interaction properties of the states like their decay widths,
masses, branching ratios, hadronic coupling constants etc. The
recently observed state $B_{J}(5840)$ has gone through various
theoretical analysis \cite{20,24} for its strong decay, but an
unique $J^{P}$ is not yet confirmed for it.

 In this paper, we confirm a particular
$J^{P}$ to the bottom state $B_{J}(5840)$ recently observed by LHCb.
On the basis of the theoretically predicted masses
Ref.\cite{20,24,34,35,36}, $B_{J}(5840)$ can be a member of the
doublets for radially excited S-wave 2S$(0^{-},1^{-})$ , or for
orbitally excited D-wave doublets 1D$(1^{-},2^{-})$ or
1D$(2^{-},3^{-})$. These six possible $J^{P}$ states are tabulated
in Table~\ref{differ} with their allowed strong decays to the ground
state bottom mesons 1S$(0^{-},1^{-})$.
\setlength{\tabcolsep}{0.09em} %
{\renewcommand{\arraystretch}{0.2}%
\begin{table*}{\normalsize
\renewcommand{\arraystretch}{1.0}
\tabcolsep 0.2cm \caption{\label{differ}  Strong decay channels for
all the six possible spin-parity $J^{P}$ values for $B_{J}(5840)$
state.} \footnotesize
\begin{tabular*}{130mm}{@{\extracolsep{\fill}}c|c|c|c|c|c|c}
\toprule
Decay Mode&2S$0^{-}$&2S$1^{-}$&1D$1^{-}$&1D$2^{-}_{3/2}$&1D$2^{-}_{5/2}$&1D$3^{-}$\\
\hline

\hline
$B^{0}\pi^{0}$&-&220.64$\widetilde{g}^{2}_{HH}$&182.58${g}^{2}_{XH}$&-&-&12.84${g}^{2}_{YH}$\\
$B^{+}\pi^{-}$&-&439.40$\widetilde{g}^{2}_{HH}$&364.14${g}^{2}_{XH}$&-&-&25.44${g}^{2}_{YH}$\\
$B^{0}\eta$&-&12.47$\widetilde{g}^{2}_{HH}$&11.60${g}^{2}_{XH}$&-&-&0.01${g}^{2}_{YH}$\\
$B_{s}K$&-&-&-&-&-&-\\
$B^{*}\pi^{0}$&523.71$\widetilde{g}^{2}_{HH}$&347.46$\widetilde{g}^{2}_{HH}$&61.63${g}^{2}_{XH}$&184.91${g}^{2}_{XH}$&16.98${g}^{2}_{YH}$&9.70${g}^{2}_{YH}$\\
$B^{*}\pi^{+}$&1040.10$\widetilde{g}^{2}_{HH}$&690.08$\widetilde{g}^{2}_{HH}$&122.46${g}^{2}_{XH}$&367.39${g}^{2}_{XH}$&33.41${g}^{2}_{YH}$&19.09${g}^{2}_{YH}$\\
$B^{*}\eta$&-&-&-&-&-&-\\
$B^{*}_{s}K$&-&-&-&-&-&-\\
Total&1563.82$\widetilde{g}^{2}_{HH}$&1710.09$\widetilde{g}^{2}_{HH}$&742.43${g}^{2}_{XH}$&552.30${g}^{2}_{XH}$&50.40${g}^{2}_{YH}$&67.09${g}^{2}_{YH}$\\
Ratio $R_{1}$&0&0.63&2.96&0&0&1.32\\
 \hline
\end{tabular*}}
\end{table*}

 To choose the best possible $J^{P}$ among these,
we study the branching ratio \begin{center} \begin{equation}BR =
R_{1} = \frac{\Gamma(B_{J}(5840)\rightarrow
B\pi)}{\Gamma(B_{J}(5840)\rightarrow B^{*}\pi)} \end{equation}
\end{center} for all these suggested $J^{P}$'s and their masses. This
ratio $R_{1}$ is effective in distinguishing these six possible
assignments, as this ratio $R_{1}$ gives result independent of the
coupling constants $\widetilde{g}_{HH}$, $g_{XH}$ and $g_{YH}$, thus
making the predictions model independent. This ratio gives different
values for all these six states, thus allowing us to notate the
proper $J^{P}$ for the bottom state $B_{J}(5840)$.

We have also plotted the graphs for the $R_{1}$ with the masses for
these $J^{P}$ states which are shown in Figure 1. It is worth
noticing that the Fig. 1(a) shows, the $R_{1}$ remains 0 for the
entire mass range which depicts that
 $B\pi$ decay mode is either suppressed or not allowed for $J^{P}$'s $2S0^{-}$,
  $1D_{3/2}2^{-}$ and $1D_{5/2}2^{-}$. The graph 1(b), 1(c) and 1(d)
 shows the variation of $R_{1}$ with the masses and give the values
 of $R_{1}$ as 0.63, 2.96 and 1.32 for
 the $J^{P}$ states $2S1^{-}$, $1D1^{-}$ and $1D3^{-}$ respectively,
 corresponding to the M(5840) =5862.90 MeV.
  The values 2.96 for $1D1^{-}$ and 1.32 for $1D3^{-}$
 point towards the dominancy of $B\pi$ mode, whereas the value 0.63
 for the $2S1^{-}$ directs the dominancy of $B^{*}\pi$ decay mode.
 The calculation of the total decay widths for all these six
classifications of $B_{J}(5840)$ requires the value of the coupling
constants $\widetilde{g}_{HH}$, ${g}_{XH}$ and ${g}_{YH}$ which are
experimentally unknown. Nevertheless, on the basis of the
theoretically available values of these couplings, following results
can be seen.

\begin{center}
\begin{figure*}

  \subfloat[ Ratio for $2S(0^{-})$, $1D(2^{-})_{3/2}$ and $1D(2^{-})_{5/2}$
  state]{\includegraphics[width=0.3\textheight]{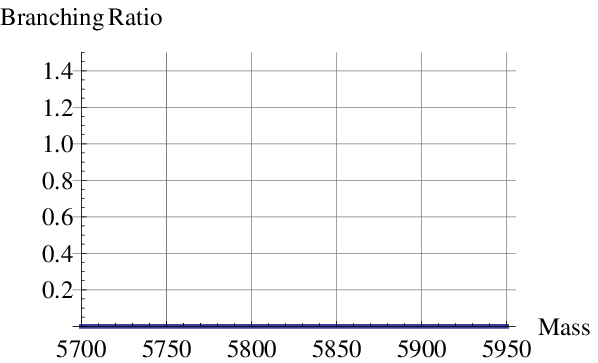}{ \label{suba}}}
\hfill
 \subfloat[ Ratio for $2S(1^{-})$
    state]{\includegraphics[width=0.3\textheight]{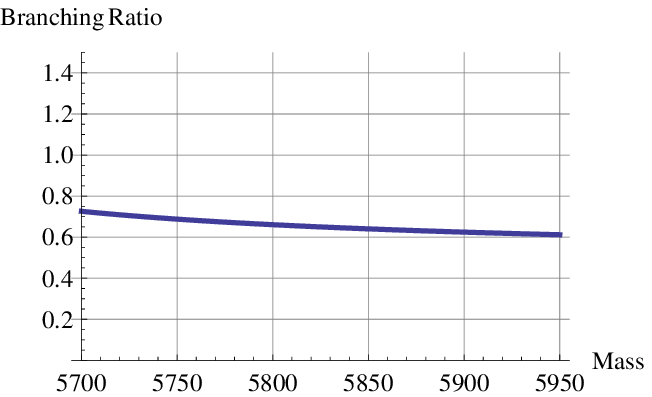}{\label{subb} }}\\
 \subfloat[ Ratio for $1D(1^{-})$
  state]{\includegraphics[width=0.3\textheight]{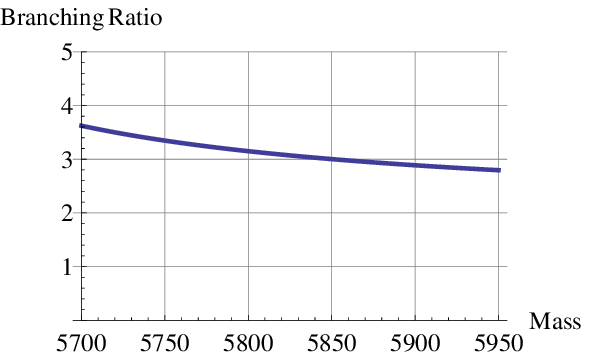}{ \label{subc}}}
\hfill
 \subfloat[Ratio for $1D(3^{-})$
    state]{\includegraphics[width=0.3\textheight]{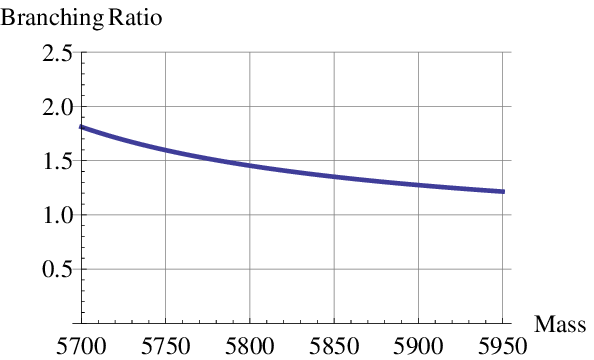}{\label{subf} }}

    \caption{Branching ratio
$\Gamma(B_{J}(5840))\rightarrow\frac{B\pi}{B^{*}\pi}$ for all six
possible $J^{P}$'s for $B_{J}(5840)$ state, where three possible
$J^{P}$'s are shown in the same fig 1(a)} \label{test}
\end{figure*}
\end{center}

\begin{itemize}
\item If $B_{J}(5840)$ is classified as the one of the member of the
doublet 2S$(0^{-},1^{-})$, then the total decay width for these
states comes out to be 150.28 MeV and 165.13 MeV respectively for
$2S0^{-}$ and $2S1^{-}$. This prediction is made using the
theoretical data $\widetilde{g}_{HH}$ = 0.31 \cite{41}. Both these
decay widths matches very well with the experimentally observed
broad decay width of 127 MeV for $B_{J}(5840)$. Since the $2S1^{-}$
state is known to be filled by the experimentally seen bottom state
$B_{J}(5970)$, and the experimentally observed decay mode
$B^{0}\pi^{+}$ is not possible for $2S0^{-}$. So the possibility of
these both $J^{P}$'s $2S0^{-}$ and $2S1^{-}$ are excluded.
\item  If $B_{J}(5840)$ is the member of the doublet
1D$(1^{-},2^{-})$ with $s_{l}^{P} = 3/2^{-}$ , then the total strong
decay width comes out to be 42.76 MeV and 31.81 MeV for $J^{P}$
$1^{-}$ and $2^{-}$ respectively. For this, ${g}_{XH}$ is taken as
0.24 which is derived using the charm state $D_{1}^{*}(2760)$
information observed by LHCb in 2016 \cite{42}. The 0 $R_{1}$ value
and the narrow decay width for state 1D$2^{-}$ also rules out this
option for the $B_{J}(5840)$.
\item The last possibility for $B_{J}(5840)$ can be the member of the
doublets 1D$(2^{-},3^{-})$. Using the available data for coupling
constant ${g}_{YH} = 0.61$ \cite{41}, the total decay widths for
$J^{P}$ states $2^{-}$ and $3^{-}$ comes out to be 18.75 MeV and
24.96 MeV respectively. Even for such high value of ${g}_{YH}$, the
decay widths are very narrow. So, the classification of
$B_{J}(5840)$ as member of 1D$(2^{-},3^{-})$ is completely ruled
out. \end{itemize}
 Thus the left out possibility of spin parity for $B_{J}(5840)$ is $1D(1^{-})_{3/2}$.
  It is interesting to
 notice that the ratio $R_{1}$ of $B_{J}(5840)$ for $J^{P}$ state $1D(1^{-})_{3/2}$ also comes
  out to be maximum with value of 2.96, thus favoring
 $1D(1^{-})_{3/2}$ as the most favorable $J^{P}$ for $B_{J}(5840)$. However,
  if we consider the fact that photon from the $B^{*}
\rightarrow
 B\gamma$ is too low in energy to be detected and $B^{*}$ mesons are
 partially reconstructed as B mesons. Then the $J^{P}$ option for
 $B_{J}(5840)$ belonging to $2S0^{-}$ cannot be fully ignored because of its large decay width.
 The identification of $B_{J}(5840)$ as $2S0^{-}$ is also supported
 by the work in \cite{9,20}. So, in future one may expect that, experimental information about the
 decay modes for $B_{J}(5840)$ broaden up, to clearly identify the
 exact $J^{P}$ for this state. In this paper, due to the only $B^{0}\pi^{+}$ experimentally
  observed decay mode, we expect $B_{J}(5840)$ to belong to $1D(1^{-})_{3/2}$ J value.

 \subsection{Prediction of Spin and Strange partners for $B_{J}(5840)$}
On the basis of the spin parity assignment of  $B_{J}(5840)$, it is
interesting to look for some features of its spin and strange
partners. As discussed, $B_{J}(5840)$ is assigned as the orbitally
excited D-wave state with $J^{P}$ as 1D$1^{-}$. The complete picture
of the partial decay widths for $B(1 ^{1}D_{2})$, $B_{s}(1
^{3}D_{1})$ and $B_{s}(1 ^{1}D_{2})$ being the spin and strange
partners of $B_{J}(5840)$ state is listed in Table~\ref{width2}.
Along with the partial decay widths, Table also shows the branching
ratios $\widehat{{\bf \Gamma}}= \frac{\Gamma}{\Gamma(B_{J}(5840)
\rightarrow B^{*+}\pi^{-})}$, $\widehat{{\bf \Gamma}}=
\frac{\Gamma}{\Gamma(B_{sJ}^{*} \rightarrow B^{*0}K^{+})}$ and
branching fractions for all the mentioned decay modes. Apart from
the decay channels mentioned in this Table, $B_{J}(5840)$ being
$1D(1^{-})$ also decays to $1 P(1^{+})$, $1 P^{'}(1^{+})$ and $1
P(2^{+})$ states along with pseudoscalar mesons $(\pi,\eta,K)$.
Since these decays occur via D-wave, so their contribution is
relatively suppressed. Here, we mentioned only the dominant decay
modes with which total decay width of $B_{J}(5840)$ comes out to be
742.43$g^{2}_{XH}$.

The information in the Table~\ref{width2} reveals that, for
$B_{J}(5840)$ state $B^{+}\pi^{-}$ and $B^{0}\pi^{0}$ are the main
decay modes as compared to the $B^{*+}\pi^{-}$ and $B^{*0}\pi^{0}$
mode. The decay width obtained in this work is finally compared with
the experimental data, and hence the coupling constant $g_{XH}$ is
obtained to be
\begin{center}
\begin{eqnarray}
 g_{XH} = 0.41\pm0.02.
 \label{eq:gxh}
\end{eqnarray}
\end{center}
This information can be a beneficial in finding total and the
partial decay widths of unobserved highly excited bottom meson
states. Theoretically, this coupling values are also obtained as
0.45 \cite{41a}, 0.53\cite{43} and 0.19\cite{21} from the charm
states $D_{sJ}(2860), D_{J}(2600)$ and bottom state $B_{J}(5960)$
assuming them to be in $1D1^{-}$ state. As the $D_{J}(2600)$ and
$B_{J}(5960)$ belongs to $2S1^{-}$, so the last two values of
coupling $g_{XH}$ = 0.53 and 0.19 predicted from $D_{J}(2600)$ and
$B_{J}(5960)$ are not useful for our study.


\setlength{\tabcolsep}{0.09em} %
{\renewcommand{\arraystretch}{0.2}%
\begin{table*}{\normalsize
\renewcommand{\arraystretch}{1.0}
\tabcolsep 0.2cm \caption{ \label{width2}  Strong decay width of
newly observed bottom mesons  $B_{J}^{*}(5840)$ and its spin and
strange partners $B(1 ^{1}D_{2})$, $B_{s}(1 ^{1}D_{2})$ and
$B_{s}^{*}(1 ^{3}D_{1})$. Ratio in 5th column represents the
$\widehat{{\bf \Gamma}}= \frac{\Gamma}{\Gamma(B_{J}^{(*)}
\rightarrow B^{*+}\pi^{-})}$ for the non-strange mesons and
$\widehat{{\bf \Gamma}}= \frac{\Gamma}{\Gamma(B_{sJ}^{*} \rightarrow
B^{*0}K^{+})}$ for the strange mesons. Branching fraction (B.R)
gives the percentage of the partial decay width with respect to the
total decay width.} \footnotesize
\begin{tabular*}{170mm}{@{\extracolsep{\fill}}ccccccc}
\toprule State&$nLs_{l}J^{P}$&Decay channel&Decay
Width/MeV&Ratio&B.F $\%$&Experimental/Theoretical value(MeV)\\
\hline $B_{J}(5840)$&1$D_{3/2}1^{-}$&$B^{*}\pi^{+}$&122.46$g^{2}_{XH}$&1&16.49&\\
&&$B^{*}\pi^{0}$&61.63$g^{2}_{XH}$&0.50&8.30&\\
&&$B^{*}\eta$&-&-&-&\\
&&$B^{*}_{s}K$&-&-&-&\\
&&$B^{0}\pi^{0}$&182.58$g^{2}_{XH}$&1.49&24.59&\\
&&$B^{+}\pi^{-}$&364.14$g^{2}_{XH}$&2.97&49.04&\\
&&$B^{0}\eta$&11.60&0.09&1.56&\\
&&$B_{s}K$&-&-&-&\\
&&Total&742.43$g^{2}_{XH}$&&&127.40 \cite{9}, 127 \cite{41b}\\
 \hline
$B_{J}(5967.20)$&1$D_{3/2}2^{-}$&$B^{*}\pi^{+}$&866.42$g^{2}_{XH}$&1&60.34&\\
&&$B^{*}\pi^{0}$&435.16$g^{2}_{XH}$&0.50&30.31&\\
&&$B^{*}\eta$&94.61$g^{2}_{XH}$&0.10&6.58&\\
&&$B^{*}_{s}K$&39.42$g^{2}_{XH}$&0.04&2.74&\\
&&Total&1435.69$g^{2}_{XH}$&&&250.69\cite{20}, 98 \cite{41b}\\
\hline
 $B_{sJ}(6083.00)$&1$D_{s(3/2)}1^{-}$&$B_{s}^{*}\pi^{0}$&175.31$g^{2}_{XH}$&0.57&5.64&\\
 &&$B_{s}^{*}\eta$&12.39$g^{2}_{XH}$&0.04&0.39&\\
 &&$B^{*0}K^{0}$&300.48$g^{2}_{XH}$&0.98&9.67&\\
 &&$B^{*-}K^{+}$&305.62$g^{2}_{XH}$&1&9.84&\\
 &&$B_{s}^{0}\pi^{0}$&488.80$g^{2}_{XH}$&1.59&15.73&\\
 &&$B_{s}\eta$&49.01$g^{2}_{XH}$&0.16&1.57&\\
 &&$B^{+}K^{-}$&894.37$g^{2}_{XH}$&2.92&28.79&\\
 &&$B^{0}K^{0}$&879.81$g^{2}_{XH}$&2.87&28.32&\\
 &&Total&3105.83$g^{2}_{XH}$&&&213.38\cite{20}, 137 \cite{41b}\\
\hline
$B_{sJ}(6057.50)$&1$D_{s(3/2)}2^{-}$&$B_{s}^{*}\pi^{0}$&436.79$g^{2}_{XH}$&0.60&23.04&\\
 &&$B_{s}^{*}\eta$&23.62$g^{2}_{XH}$&0.03&1.24&\\
  &&$B^{*-}K^{+}$&724.42$g^{2}_{XH}$&1&38.21&\\
&&$B^{*0}K^{0}$&710.60$g^{2}_{XH}$&0.98&37.48&\\
&&Total&1895.45$g^{2}_{XH}$&&&198.64\cite{20}, 89 \cite{41b}\\
\hline
\end{tabular*}}
\end{table*}

Theoretically, mass of the spin partner of $B_{J}(5840)$ i.e. $B(1
^{1}D_{2})$ is predicted to be $5967.20\pm30$ MeV in
Ref.\cite{20,24,34,35,36}. Column 5 of the Table gives the ratio of
the partial decay widths for $B(1 ^{1}D_{2})$ with respect to its
partial decay width $B^{*-}\pi^{+}$.
 Apart from the decay channels listed in this Table, $B(1 ^{1}D_{2})$ also decays to P-wave
bottom meson states $1 P(0^{+})$, $1 P(1^{+})$, $1 P^{'}(1^{+})$ and
$1 P(2^{+})$ which occurs via D-wave, and thus due to the small
phase space, these decay modes are suppressed when compared with
decays to ground state S-wave mesons and hence are not shown in
Table~\ref{width2}. From the listed decay channels, $B^{*-}\pi^{+}$
comes out to be the dominant decay mode for $B(1 ^{1}D_{2})$ with
branching fraction $60.34\%$. Decay width calculated in Ref.
\cite{41b}, also shows $B^{*}\pi$ to be the dominant decay mode.
Hence, the decay mode $B^{*-}\pi^{+}$ can be a motivation for the
experimental search for the missing bottom state $B(1 ^{1}D_{2})$ in
future. Using the value of the coupling constant $g_{XH}$ obtained
from equation \ref{eq:gxh}, the total decay width of the bottom
state $B(1 ^{1}D_{2})$ is obtained as 241.33 MeV. This decay width
value is in same range as given in Ref. \cite{20} with 3.87 \%
deviation.
\\ Masses for the strange partners of these bottom
states are taken as $6083.06$ MeV and $6057.50$ MeV from the
theoretical work \cite{20,24,34,35,36}. Referring to the branching
fractions in the Table~\ref{width2}, $B^{+}K^{0}$ and $B^{*-}K^{+}$
seems to be the dominant decay modes with contribution $28.79\%$ and
$38.21\%$ for the bottom strange states $B_{s1}^{*}$ and $B_{s2}$
respectively, which are comparable with the dominant modes seen in
Ref. \cite{41b}. These strange states also allow decays to P-wave
bottom meson states, but are relatively suppressed. Hence the total
decay width for these strange state comes out to be
\begin{center}
\begin{eqnarray}
\Gamma(B_{s1}^{*}) = 522.09 MeV.\\
 \Gamma(B_{s2}) = 318.62 MeV.
\end{eqnarray}
\end{center}

The results conclude $B_{s1}^{*}$ to be a broader state as compared
to its spin partner $B_{s2}$. Moreover, if we use the coupling
$g_{XH} =  0.45$ obtained in Ref.\cite{41a}, the decay widths for
states $B(1^{1}D_{2}), B_{s}(1^{3}D_{1})$
 and $B_{s}(1^{1}D_{2})$ deviate from our results by $16\%$.

\subsection{Analysis for bottom states $B_{1}(5721)$, $B_{2}^{*}(5747)$, $B_{1s}(5830)$ and
$B_{2s}^{*}(5840)$} We have also analyzed the bottom states
$B_{1}(5721)$, $B_{2}^{*}(5747)$, $B_{1s}(5830)$ and
$B_{2s}^{*}(5840)$ for their $J^{P}$'s. On the basis of their
available theoretical and experimental information, the states
$B_{1}(5721)$, $B_{2}^{*}(5747)$, $B_{1s}(5830)$ and
$B_{2s}^{*}(5840)$ are identified as the P- wave bottom mesons with
$j_{l}$ = 3/2.
\begin{center}
\begin{eqnarray}
\label{eq:lagrangian} (B_{1}(5721), B_{2}^{*}(5747)) = (1^{+},
2^{+})_{3/2} \text{with }n = 1, L = 1. \\
(B_{1s}(5830), B_{2s}^{*}(5840)) = (1^{+}, 2^{+})_{3/2} \text{with
}n = 1, L = 1 .
\end{eqnarray}
\end{center}
We study their strong decay widths using the equations 14-16 and
calculate the various branching ratios involved. The numerical value
of the partial decay widths for the bottom states $B_{1}(5721)$,
$B_{2}^{*}(5747)$, $B_{1s}(5830)$ and $B_{2s}^{*}(5840)$ are given
in Table~\ref{width1}. The obtained decay widths are then compared
with the experimental data to obtain the strong coupling constant
$g_{TH}$. Since the strange states $B_{1s}(5830)$ and
$B_{2s}^{*}(5840)$
 are very narrow, we exclude them to calculate the coupling constant $g_{TH}$. $g_{TH}$
comes out to be $0.50\pm0.01$ and $0.37\pm0.01$ for bottom states
$B_{1}(5721)$ and $B_{2}^{*}(5747)$ respectively. This is consistent
with the theoretical values of $g_{TH}$ in Ref.\cite{41}, \cite{21},
\cite{40} obtained from charm mesons. Here, the consistency in the
hadronic coupling constant $g_{TH}$ beautifully describes the heavy
quark symmetry between the charm and the bottom mesons.
\setlength{\tabcolsep}{0.09em} %
{\renewcommand{\arraystretch}{0.2}%
\begin{table*}{\normalsize
\renewcommand{\arraystretch}{1.0}
\tabcolsep 0.2cm \caption{ \label{width1}  Strong decay width of
newly observed bottom mesons $B_{1}(5721)$ and $B_{2}^{*}(5747)$ and
their strange partners $B_{1s}(5830)$ and $B_{2s}^{*}(5840)$. Ratio
in 5th column represents the $\widehat{{\bf \Gamma}}=
\frac{\Gamma}{\Gamma(B_{J}^{(*)} \rightarrow B^{*+}\pi^{-})}$ for
the non-strange mesons and $\widehat{{\bf \Gamma}}=
\frac{\Gamma}{\Gamma(B_{sJ}^{*} \rightarrow B^{*0}K^{+})}$ for the
strange mesons. Branching fraction (B.F) gives the percentage of the
partial decay width with respect to the total decay width.}
\footnotesize
\begin{tabular*}{170mm}{@{\extracolsep{\fill}}ccccccc}
\toprule State&$nLs_{l}J^{P}$&Decay channel&Decay
Width(MeV)&Ratio&B.F $\%$&Experimental/Theoretical value(MeV)\\
\hline $B_{1}(5721)$&1$P_{3/2}1^{+}$&$B^{*}\pi^{+}$&74.83$g^{2}_{TH}$&1&66.16&\\
&&$B^{*}\pi^{0}$&38.25$g^{2}_{TH}$&0.51&33.82&\\
&&Total&113.09$g^{2}_{TH}$&&&30.1 \cite{9}, 55 \cite{41b}\\
 \hline
$B_{2}^{*}(5747)$&1$P_{3/2}2^{+}$&$B^{*}\pi^{+}$&52.47$g^{2}_{TH}$&1&30.65&\\
&&$B^{*}\pi^{0}$&26.78$g^{2}_{TH}$&0.51&15.64&\\
&&$B^{0}\pi^{0}$&30.91$g^{2}_{TH}$&0.58&18.05&\\
&&$B^{+}\pi^{-}$&61.02$g^{2}_{TH}$&1.16&35.64&\\
&&Total&171.18$g^{2}_{TH}$&&&24.5\cite{9}, 7 \cite{41b}\\
\hline
$B_{1s}(5830)$&1$P_{s3/2}1^{+}$&$B^{*}_{s}\pi^{0}$&44.45$g^{2}_{TH}$&-&100&\\
&&$B^{*+}K^{0}$&$\sim \mathcal{O}(10^{-10})$$g^{2}_{TH}$&-&-&\\
&&$B^{*-}K^{+}$&$\sim \mathcal{O}(10^{-10})$$g^{2}_{TH}$&-&-&\\
&&Total&44.45$g^{2}_{TH}$&&&0.5 \cite{38}, 30 \cite{41b}\\
\hline
$B_{2s}^{*}(5840)$&1$P_{s3/2}2^{+}$&$B^{*}_{s}\pi^{0}$&30.41$g^{2}_{TH}$&87.38&40.27\\
&&$B^{*+}K^{0}$&$0.20g^{2}_{TH}$&0.59&0.27\\
&&$B^{*-}K^{+}$&$0.34g^{2}_{TH}$&1&0.46\\
&&$B^{+}K^{-}$&3.97$g^{2}_{TH}$&11.40&5.25\\
&&$B^{0}K^{0}$&4.64$g^{2}_{TH}$&13.33&6.14\\
&&$B_{s}\pi^{0}$&35.92$g^{2}_{TH}$&103.21&47.53\\
&&Total&75.57$g^{2}_{TH}$&&&1.40\cite{38}, 1 \cite{41b}\\
 \hline
\end{tabular*}}
\end{table*}
 We also obtained the ratios $R_{2}$, $R_{3}$ and
$R_{4}$ as
\begin{gather}
R_{2} =
\frac{\Gamma(B_{1}(5721))}{\Gamma(B_{1}(5721))+\Gamma(B_{2}^{*}(5747))} = 0.60\\
R_{3} = \frac{\Gamma(B_{2}^{*}\rightarrow
B^{*}\pi)}{(\Gamma(B_{2}^{*}\rightarrow
B^{*}\pi)+\Gamma(B_{2}^{*}\rightarrow B\pi))} = 0.46\\
R_{4} = B_{2}^{*}(5747)\rightarrow\frac{B^{*+}\pi^{-}}{B^{+}\pi^{-}}
=0.85 \thinmuskip \thinmuskip \qquad \qquad
\end{gather}
which are consistent with their experimental value $R_{2} =
0.47\pm0.06$ and $R_{3} = 0.47\pm0.09$ observed by D0 collaboration
\cite{10} and $R_{4} = 0.71\pm0.14$ measured by LHCb \cite{9}.
Table~\ref{width1} also shows the decay widths of the strange bottom
states $B_{1s}(5830)$ and $B_{2s}^{*}(5840)$. The negligible values
of the decay widths (of order of $10^{-10}$ MeV) for $B_{1s}(5830)$
state decaying to $B^{*+}K^{-}$ and
 $B^{*-}K^{0}$ are consistent with its very small decay width 0.5 MeV measured by CDF collaboration \cite{38} in 2014.
Table~\ref{width1} reveals that $B^{*+}\pi^{-}$ and $B^{+}\pi^{-}$
are main decay modes for $B_{1}(5721)$ and $B_{2}^{*}(5747)$ with
branching fraction $66.16\%$ and $35.64\%$ respectively. Similarly
$B^{*}_{s}\pi^{0}$ and $B_{s}\pi^{0}$ are observed to be the
dominating decay modes for their strange partners $B_{1s}(5830)$ and
$B_{2s}^{*}(5840)$ respectively.
 \subsection{Prediction of Spin and Strange partners for $B_{J}(5970)$}
Now, we will proceed in the similar manner to study the spin and
strange partners for bottom state $B_{J}(5970)$. As we have
discussed, $B_{J}(5970)$ is fitted to be the radially excited state
with $J^{P}$ $1^{-}$. Table~\ref{width3} shows the partial decay
widths for $B_{J}(5970)$ along with its spin and strange partners
$B(2 ^{1}S_{0})$, $B_{s}(2 ^{3}S_{1})$ and
  $B_{s}(2 ^{1}S_{0})$. Along with the partial decay widths, Table
  also shows the branching ratio $\widehat{{\bf \Gamma}}= \frac{\Gamma}{\Gamma(B_{J}^{(*)}
\rightarrow B^{*+}\pi^{-})}$ and $\widehat{{\bf \Gamma}}=
\frac{\Gamma}{\Gamma(B_{sJ}^{*} \rightarrow B^{*0}K^{+})}$ for the
non-strange and strange states $B(2 ^{1}S_{0})$,$B(2 ^{3}S_{1})$,
$B_{s}(2 ^{3}S_{1})$ and $B_{s}(2 ^{1}S_{0})$ respectively.
\setlength{\tabcolsep}{0.09em} %
{\renewcommand{\arraystretch}{0.2}%
\begin{table*}{\normalsize
\renewcommand{\arraystretch}{1.0}
\tabcolsep 0.2cm \caption { \label{width3}  Strong decay width of
bottom meson $B_{J}(5970)$ with its spin and strange partners $B(2
^{1}S_{0})$, $B_{s}(2 ^{1}S_{0})$ and $B_{s}(2 ^{3}S_{1})$. Ratio in
5th column represents the $\widehat{{\bf \Gamma}}=
\frac{\Gamma}{\Gamma(B_{J}^{(*)} \rightarrow B^{*+}\pi^{-})}$ for
the non-strange mesons and $\widehat{{\bf \Gamma}}=
\frac{\Gamma}{\Gamma(B_{sJ}^{*} \rightarrow B^{*0}K^{+})}$ for the
strange mesons. Branching fraction (B.F) gives the percentage of the
partial decay width with respect to the total decay width.}
\footnotesize
\begin{tabular*}{170mm}{@{\extracolsep{\fill}}ccccccc}
\toprule State&$nLs_{l}J^{P}$&Decay channel&Decay
Width(MeV)&Ratio&B.F $\%$ &Experimental/Theoretical value(MeV)\\
\hline
$B_{0}(5881)$& 2$^{1}S_{0}0^{-}$&$B^{*}\pi^{+}$&1148.08$\widetilde{g}^{2}_{HH}$&1&66.48&\\
&&$B^{*}\pi^{0}$&577.80$\widetilde{g}^{2}_{HH}$&0.50&33.27&\\
&&$B^{*}\eta$&1.00$\widetilde{g}^{2}_{HH}$&0.00&0.05&\\
&&$B^{*}_{s}K$&-&-&-&\\
&&Total&1726.89$\widetilde{g}^{2}_{HH}$&&&91 \cite{41b}\\
\hline
$B_{J}(5970)$& 2$^{3}S_{1}1^{-}$&$B^{*}\pi^{+}$&1178.23$\widetilde{g}^{2}_{HH}$&1&36.30&\\
&&$B^{*}\pi^{0}$&591.95$\widetilde{g}^{2}_{HH}$&0.50&18.23&\\
&&$B^{*}\eta$&122.22$\widetilde{g}^{2}_{HH}$&0.10&3.76&\\
&&$B^{*}_{s}K$&69.94$\widetilde{g}^{2}_{HH}$&0.05&2.15&\\
&&$B^{0}\pi^{0}$&359.11$\widetilde{g}^{2}_{HH}$&0.30&11.06&\\
&&$B^{+}\pi^{-}$&716.21$\widetilde{g}^{2}_{HH}$&0.60&22.06&\\
&&$B^{0}\eta$&113.37$\widetilde{g}^{2}_{HH}$&0.09&3.49&\\
&&$B_{s}K$&94.43$\widetilde{g}^{2}_{HH}$&0.08&2.90&\\
&&Total&3245.49$\widetilde{g}^{2}_{HH}$&&&82.30 \cite{9}, 107 \cite{41b}\\
\hline

$B_{s0}(5976.0)$&$(2 ^{1}S_{0})0^{-} $&$B^{*0}K^{0}$&521.96$\widetilde{g}^{2}_{HH}$&0.96&31.38&\\
&&$B^{*+}K^{-}$&539.41$\widetilde{g}^{2}_{HH}$&1&32.43&\\
&&$B^{*}_{s}\pi^{0}$&593.72$\widetilde{g}^{2}_{HH}$&1.10&35.70&\\
&&$B^{*}_{s}\eta$&7.85$\widetilde{g}^{2}_{HH}$&0.01&0.47&\\
&&Total&1662.97$\widetilde{g}^{2}_{HH}$&&&75.80\cite{23}, 106 \cite{41b}\\

\hline $B^{*}_{s}$(6007.8)&$ 2 ^{3}S_{1}1^{-}$&$B^{0}K^{0}$&342.71$\widetilde{g}^{2}_{HH}$&0.97&15.87&\\
&&$B^{+}K^{-}$&350.66$\widetilde{g}^{2}_{HH}$&1&16.24&\\
&&$B_{s}\pi^{0}$&292.67$\widetilde{g}^{2}_{HH}$&0.83&13.55&\\
&&$B_{s}\eta$&58.29$\widetilde{g}^{2}_{HH}$&0.16&2.70&\\
&&$B^{*0}K^{0}$&474.96$\widetilde{g}^{2}_{HH}$&1.35&22.00&\\
&&$B^{*+}K^{-}$&486.47$\widetilde{g}^{2}_{HH}$&1.38&22.53&\\
&&$B^{*}_{s}\pi^{0}$&466.53$\widetilde{g}^{2}_{HH}$&1.33&21.61&\\
&&$B^{*}_{s}\eta$&36.98$\widetilde{g}^{2}_{HH}$&0.10&1.71&\\
&&Total&2158.65$\widetilde{g}^{2}_{HH}$&&&114.0\cite{23}, 127 \cite{41b}\\
 \hline
\end{tabular*}}
\end{table*}

  From the experimental decay widths of $B_{J}(5970)$, we obtain the strong coupling constant
  $\widetilde{g}_{HH}$ as
  \begin{center}
\begin{eqnarray}
  \widetilde{g}_{HH} = 0.15 \pm 0.01.
\label{eq:24}
\end{eqnarray}
\end{center}
  The error in the value of coupling comes from the statistical error in
   experimental mass and decay width values of these bottom states.
  Using HQET, this coupling constant $\widetilde{g}_{HH}$ is also
  predicted as 0.14\cite{21}, 0.31\cite{41}, 0.28\cite{17} and 0.40\cite{43}. The first value
  is obtained from bottom state $B_{J}(5960)$
   and other three values are obtained from the charm state sector
   by assuming the charm states to be in $2S0^{-}$ state.

  From the listed decay channels mentioned in Table~\ref{width3},
$B^{*-}\pi^{+}$ comes out to be the dominant decay mode for
$B_{J}(5970)$ and its spin partner $B(2 ^{1}S_{0})$ with branching
fraction $36.30\%$ and $66.48\%$ respectively.

 Apart from the decay channels listed in
Table~\ref{width3}, we also find its partial decays to $1 P(0^{+})$,
$1 P(2^{+})$, $1 D(1^{-})$ and $1 D(3^{-})$ states, but due to the
small phase space, these decay modes are suppressed and are not
considered in this work. And for their strange partners $B_{s0}$ and
$B_{s1}^{*}$, we observe, $B^{*+}K^{-}$ and $B^{*}_{s}\pi^{0}$ as
the dominant decay modes for $B_{s1}^{*}$ and $B_{s0}$ bottom states
respectively. Thus these decay modes are suitable for the
experimental search for these missing radially excited strange
bottom mesons $B_{s1}^{*}$ and $B_{s0}$. Using the result in
equation \ref{eq:24}, their total decay widths corresponding to the
mass $M(B_{0})$ = 5881.00 MeV, $M(B_{s1}^{*})$ = 6007.80 MeV and
$M(B_{s0})$ = 5976.00 MeV \cite{20,24,34,35,36} are obtained as
\begin{gather}
\Gamma(B_{0}) = 38.85 MeV\\
\Gamma(B_{s1}^{*}) = 37.41 MeV \\
\Gamma(B_{s0}) = 48.56 MeV
\end{gather}
This shows that the strange partners follow the same pattern as the
non-strange bottom states i.e. $B_{s1}^{*}$ state is seen to be
broader as compared to its spin partner $B_{s0}$.
 From the other theoretical available coupling values of
$\widetilde{g}_{HH}$, the upper most theoretical value predicted
from charm states is 0.40\cite{43}. The results for decay widths
obtained using this higher value are very large $(\simeq 254 \pm
35)$ MeV from the values obtained in our result. However, if we use
the coupling value $\widetilde{g}_{HH}$ = 0.14 \cite{21}obtained
from bottom sector, it give decay widths as 33.84 MeV, 32.59 MeV,
42.30 MeV for states $B(2^{1}S_{0}), B_{s}(2^{1}S_{0}),
B_{s}(2^{3}S_{1})$, which deviates by $11\%$ from our results. The
results for dominating decay modes for all these four states are
same as observed in Ref. \cite{41b}. If we look at the leading order
terms of coupling constants, it will remain same for both charm and
bottom sectors. It may vary if we go for corrections upto $1/m_{Q}$
order. So using the coupling constant values obtained from
experimental charm states to theoretically predict the information
for bottom states may change the actual results. Moreover, one can
also extend the work by studying the decays decaying to ground state
 through vector mesons with $J^{P} = 1^{-}$ \cite{42}.

  \section{Conclusion}
In the present article, we have used the  Heavy Quark Effective
Theory of Ref.\cite{17} to investigate the recently observed bottom
mesons, $B_J(5840)$, $B_2^{*}(5747)$,$B_J^{}(5840)$,
$B_J^{*}(5960)$, $B_{s1}^{*}(5830)$ and $B_{s2}^{*}(5840)$ by
calculating the $B/B^{*}$ -light pseudoscalar meson decay widths. We
also calculate the strong decay widths for the experimentally
unobserved but theoretically predicted states $B(2 ^{1}S_{0})$,
$B_{s}(2 ^{3}S_{1})$, $B_{s}(2 ^{1}S_{0})$, $B(1 ^{1}D_{2})$,
$B_{s}(1 ^{3}D_{1})$ and $B_{s}(1 ^{1}D_{2})$.
In particular, we have identified the six possible
spin parity assignments for $B_{J}(5840)$ state, observed by the
LHCb in 2015 \cite{9}. We have analyzed the total decay widths and
branching ratio ($R_{1}$) $\frac{B\pi}{B^{*}\pi}$ for all these six
assignments in Table~\ref{differ} and concluded that the only
favorable J value for $B_{J}(5840)$ state is $1D1^{-}$. This ratio
has very different values for $B_{J}(5840)$ belonging to these two
$J^{P}$'s, so experimental measurement of such a branching ratio in
the future will be very helpful in clearly identifying one of them
to be the most favorable $J^{P}$ for $B_{J}(5840)$.
\\ We have also obtained coupling
constant $g_{XH}$, $\widetilde{g}_{HH}$ and $g_{TH}$
 governing the strong decays of bottom states to the light pseudo-scalar
mesons. These obtained couplings allowed us to compute the strong
decay widths of the above mentioned experimentally missing bottom
states. Along with this, we examine the recently observed bottom
state $B_{J}(5721)$ and $B_{2}^{*}(5747)$ and their strange partners
$B_{sJ}(5830)$ and $B_{2s}^{*}(5840)$ for their $J^{P}$'s as
$1P_{3/2}1^{+}$ and $1P_{3/2}2^{+}$ respectively. Thus, these
predictions have opened a window to investigate the higher
excitations of bottom mesons at the LHCb, D0, CDF.

\end{document}